\documentclass[12pt,a4paper]{iopart}

\usepackage{graphicx}
\usepackage{amsfonts}
\usepackage{color}
\newcommand{\be}{\begin{eqnarray}}
\newcommand{\ee}{\end{eqnarray}}
\newcommand{\bw}{\begin{widetext}}
\newcommand{\ew}{\end{widetext}}

\def\ket#1{{|#1\rangle}}
\def\bra#1{{\langle #1 |}}
\def\br{{\vecsymb{r}}}

\def\bq{{\vecsymb{q}}}

\def\vecsymb#1{{\vec{#1}}}

\begin{document}

\title{Quantum Hall Valley Nematics}

\author{S. A. Parameswaran}
\address{Rudolf Peierls Centre for Theoretical Physics, Clarendon Laboratory, University of Oxford, Oxford OX1 3PU, UK}
\ead{sid.parameswaran@physics.ox.ac.uk}

\author{B. E. Feldman}
\address{Department of Physics, Stanford University, Stanford, California 94305, USA}
\ead{bef@stanford.edu}

\vspace{5pt}
\begin{indented}
\item[]September 2018
\end{indented}

\begin{abstract}
Two-dimensional electron gases in strong magnetic fields provide a canonical platform for realizing a variety of electronic ordering phenomena. Here we review the physics of one intriguing class of interaction-driven quantum Hall states: quantum Hall valley nematics. These phases of matter emerge when the formation of a topologically insulating quantum Hall state is accompanied by the spontaneous breaking of a point-group symmetry that combines a spatial rotation with a permutation of valley indices. The resulting orientational order is particularly sensitive to quenched disorder, while quantum Hall physics links charge conduction to topological defects. We discuss how these combine to yield a rich phase structure, and their implications for transport and spectroscopy measurements. In parallel, we discuss relevant experimental systems. We close with an outlook on future directions.
\end{abstract}

%
%
%
%
%

\section{Introduction}
Two-dimensional electron gases (2DEGs) in high magnetic fields are an enduring source of new physical phenomena, and they form the setting for many unconventional strongly correlated electronic phases~\cite{Pinzuk95}. Prominent among these are the integer and fractional quantized Hall (QH) states, which remain among the best-studied examples of topological phases of matter~\cite{Wen:1990p1}. Beyond the myriad incompressible QH states, 2DEGs also host several compressible phases, ranging from those with broken symmetries such as the low-density Wigner crystal and bubble- and stripe-ordered phases in high Landau levels~\cite{PhysRevLett.76.499}, to the enigmatic $\nu=1/2$ composite fermion Fermi liquid in the half-filled Landau level~\cite{Halperin:1993p144}.  Additional degrees of freedom, such as those associated with electron spin~\cite{SondhiSkyrmion} or ``pseudospin''  (e.g., corresponding to different heterostructure layers~\cite{Moon:1995p1,Jungwirth2000}, subbands~\cite{Piazza1999} or  Landau level indices~\cite{Barlas2008,Hunt2017}), lead to further richness. Such multicomponent QH systems can support incompressible states characterized by both topological structure and broken symmetry. Collectively termed {\it quantum Hall ferromagnets} (QHFMs),  these states of matter host a variety of unique phenomena --- such as electrically charged topological defects~\cite{SondhiSkyrmion}, unusual collective modes~\cite{Rasolt:1986p1}, and Josephson-like effects~\cite{Moon:1995p1,Eisenstein:2004aw} --- that stem from the interplay of topology with broken symmetry, which links the charge and spin/pseudospin degrees of freedom.

An especially rich class of QHFMs~\cite{Rasolt:1986p1,abanin_nematic_2010} emerges when the symmetry in question relates two or more degenerate conduction-band minima, or ``valleys'', of the 2DEG. A valley symmetry operation permutes valley indices while simultaneously transforming crystalline symmetry axes via a point-group operation. This intertwining of internal and spatial symmetries means that a valley degree of freedom generically {\it cannot} be viewed as purely internal: except in special cases, breaking a valley symmetry also breaks the (discrete) rotational symmetry of the underlying crystalline host material. In other words, valley QHFMs generically  have {\it nematic} order and are extremely sensitive to orientational symmetry-breaking terms in the Hamiltonian~\cite{abanin_nematic_2010}. 

We note at the very outset that {\it valley} nematic order is quite distinct from two other `quantum hall nematics' that are often encountered in the literature~\cite{NematicAnnuRev}. The first of these is  the compressible nematic  
metal that emerges near half-filling in high Landau levels~\cite{DU1999389,PhysRevLett.82.394,PhysRevLett.84.1982}; the second is a  fractional nematic QH liquid~\cite{Xia:2011rt,PhysRevB.82.085102,PhysRevB.84.195124,PhysRevB.88.125137,PhysRevX.4.041050} obtained when strong electron correlations lead to spontaneous breaking of {\it continuous} orientational symmetry in a single-valley 2DEG. In contrast to the first example, the quantum Hall valley nematics  we discuss here are gapped in the bulk and have a robust QH response (at least in the absence of domain formation). In contrast to the second, valley nematics emerge only in situations where orientational symmetry is discrete from the outset --- as evinced by the presence of multiple valleys related by discrete rotations.

Such valley nematic QH phases are the subject of this review. Excitingly, they have robust signatures visible to various experimental probes and in several different material systems: 2DEGs in conventional semiconductors with highly anisotropic valley dispersion ---  such as AlAs~\cite{Shayegan2006} and Si~\cite{eng_integer_2007} ---  exhibit  transport phenomena consistent with nematic order. Most strikingly, recent scanning tunneling microscopy (STM) experiments on the surface of elemental bismuth (Bi) have visualized a nematic electronic liquid and its associated domain structure directly in real space~\cite{Feldman2016,Randeria2018}, and provide a platform to investigate domain-wall mediated transport, key predictions of the theory of nematic QHFMs. We discuss these and other experimental developments below. Since a new generation of experiments has started to probe the interplay of interactions with the valley degree of freedom in graphene~\cite{Barlas2012}, transition metal dichalcogenides \cite{Xu2014,Xu2016}, and van der Waals heterostructures~\cite{Rivera2016}, we anticipate there will be many new experiments where these ideas will prove relevant.

This review is organized as follows. In Section~\ref{sec:VQHFM-th} we introduce a microscopic model for a two-valley nematic QHFM (relevant to AlAs) that will serve as a workhorse example throughout the manuscript. In Section~\ref{sec:EFT} we briefly summarize the key aspects of an effective-field theory description of the nematic QHFM before moving to discuss the role of quenched disorder in Section~\ref{sec:disorder} and the properties of topological defects and their role in transport phenomena in Section~\ref{sec:topodeftransport}. Armed with this basic framework, in Section~\ref{sec:complexity} we further generalize the theory to treat more complex situations that can arise due to continuous symmetries as well as new effects owing to the inclusion of spin-orbit coupling and  inversion symmetry breaking. In Section~\ref{sec:summary} we close with an outlook and a discussion of two possible extensions of these ideas: to fractional QHE states, and to three-dimensional multi-valley systems. Throughout, we examine various different experimental systems as illustrative examples as we develop each aspect of the theory.

\section{Valley Ferromagnets: Microscopic Theory \label{sec:VQHFM-th}}
\begin{figure}
\includegraphics[width=.75\columnwidth]{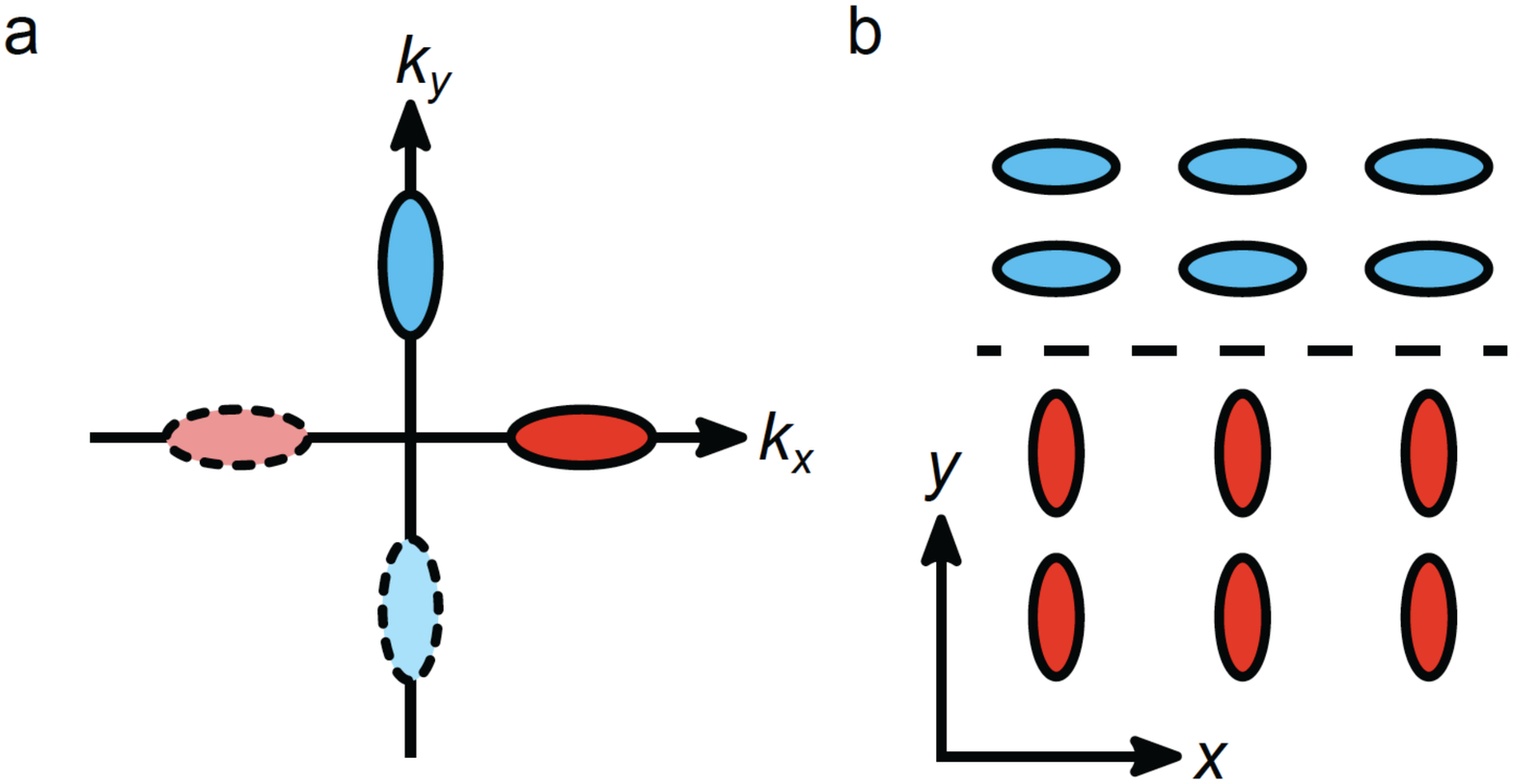}
\caption{\label{fig:AnisotropicValleys}(a) Model band structure for the $\mathcal{N}=2$ case consisting of a pair of valleys centered at the X points of the BZ with anisotropy axes rotated by $\pi/2$ with respect to each other. (b) Sketch of two possible nematic QHFM states.}
\end{figure}

The minimal microscopic model of a two-valley 2DEG, that will underpin much of our discussion, is described by the single-particle Hamiltonian \be
H = \sum_{\kappa=1,2} H_\kappa,\,\,\,\,\,\,\,\,\,\,\, H_\kappa = \sum_{i=x,y} \frac{\left(p_i - K_{\kappa,i} + {e} A_i\right)^2}{2m_{\kappa,i}},
\ee 
where $\vec{K}_1 = (K_0,0)$, $\vec{K}_2 = (0,K_0)$ are the positions of the two valleys in the Brillouin zone (BZ), and we have $m_{1,x}/m_{1,y} = \lambda^2 = m_{2,y}/m_{2,x}$ (Fig. \ref{fig:AnisotropicValleys}). The specific orientation of the valleys and their anisotropy are chosen to match those in AlAs wide quantum wells~\cite{Shayegan2006}; in accord with this, we take $K_0$ to be one half a reciprocal lattice vector, corresponding to valleys centered at the edges of the BZ, and $\lambda^2 = 5$. Other choices of $K_0$ will lead to valleys centered inside the BZ, and then on symmetry grounds we would expect four or six such valleys, as in the Si/Bi examples discussed below. It is convenient to work in Landau gauge, $\vec{A} = (0, -Bx)$, so that we may label eigenstates by their momentum $p_y$, measured with respect to $\vec{K}_{\kappa}$. Then,   the single-particle states in the $n^{\mathrm{th}}$ Landau level (LL) at energy $E_n = (n+1/2)\hbar \omega_c$ take the form 
\be
\psi^{(n)}_{\kappa, X}({\br}) = \frac{e^{iXy/\ell_B^2}}{\sqrt{L_y \ell_B}} 
\phi_n(x-X; u_\kappa)e^{i \vec{K}_{\kappa}\cdot\vec{r}} \label{eq:LLwav}
\ee
where $u_1 = 1/u_2 = \lambda$, $\ell_B = (\hbar/eB)^{1/2}$, and  $\phi_n(x-X; u_\kappa)$ is the $n^{\mathrm{th}}$ eigenfunction of a 1D simple harmonic oscillator with oscillator length $\ell_B/u_\kappa$ and its potential minimum at the ``guiding center'' $X$. We ignore spin, and assume the 2DEG is fully spin-polarized; below we will comment on situations where this must be revisited. The degeneracy of a LL is given by the number of flux quanta threading a sample of area $A$, $N_\Phi = BA/\Phi_0$, where $\Phi_0 = h/e$ is the quantum of flux.  A central quantity for QH physics is the LL filling factor,  $\nu$, that measures the number of electrons per available electronic state: $\nu =  N_e/N_\Phi$, where $N_e$ is the number of electrons.
 
\begin{figure}
\includegraphics[width=.75\columnwidth]{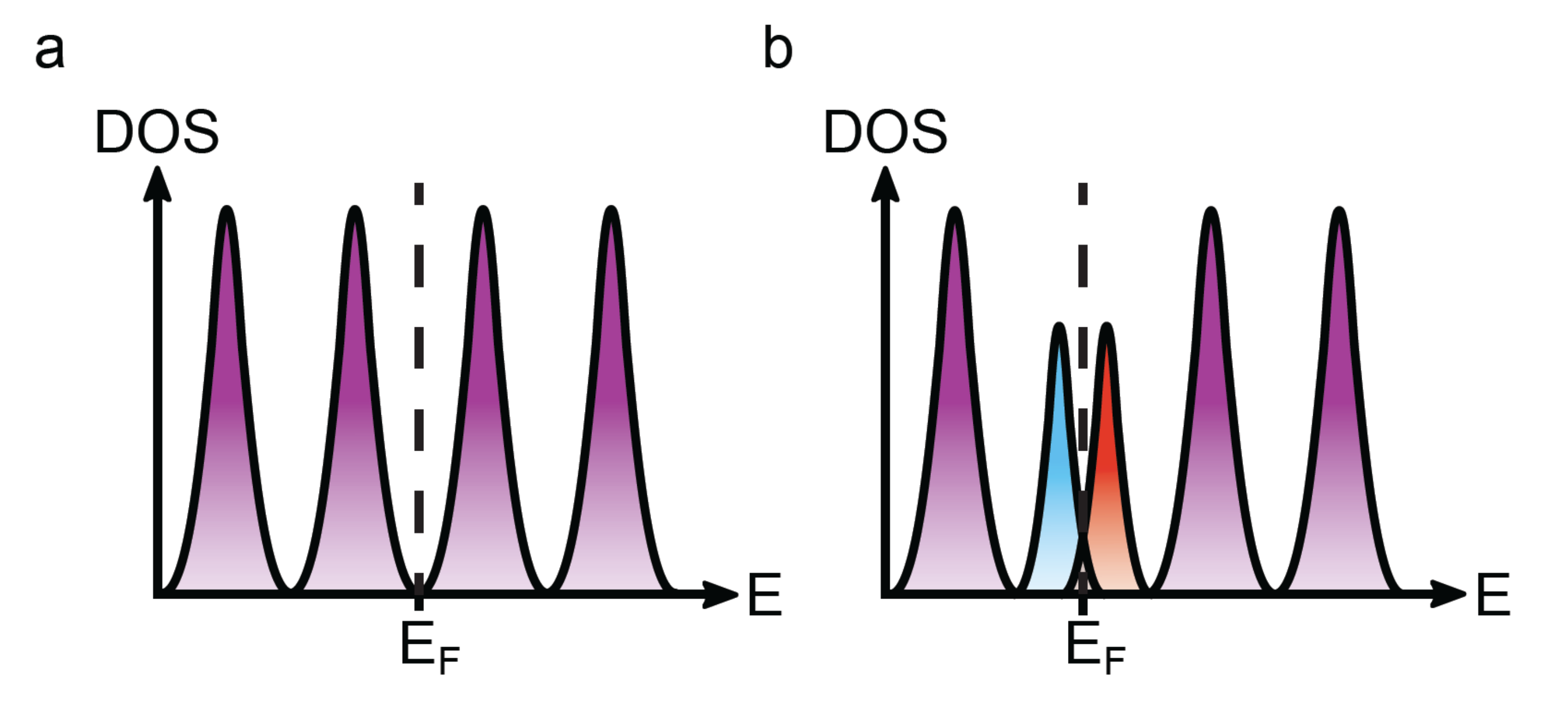}
\caption{\label{fig:ExchangeSplitting} Disorder-broadened Landau levels. (a.) In the absence of interactions, QH states only emerge when the Fermi energy is such that $\nu =  \mathcal{N}p$ Landau levels are filled; here we assume $\mathcal{N}=2$ (b.) Exchange interactions split the degeneracy, so that LLs of the two valleys are now separated. Note that even if the disorder-broadening is comparable to the exchange gap,  so that the valley-split LLs overlap in the presence of disorder, a robust QH plateau is observed as long as states at the Fermi energy are localized. As the system is tuned across the center of the QH plateau at $\nu=p=1$, the `valley character' of the localized states changes, leading to a change in the longitudinal transport anisotropy.
}
\end{figure}

 In the absence of internal degrees of freedom, if $\nu$ is an integer then the chemical potential lies in a gap between LLs, and the emergence of a quantized Hall conductance can be explained within a non-interacting picture, though disorder is usually invoked to explain the physics of Hall plateaux~\cite{RevModPhys.67.357}. Interactions are then only relevant to explaining the FQHE. However, when there is an $\mathcal{N}$-fold internal degeneracy, at integer fillings that are {\it not} multiples of $\mathcal{N}$, interactions can lift the degeneracy and produce additional gapped (integer) QH states (Fig. \ref{fig:ExchangeSplitting}). In the case at hand, $\mathcal{N}=2$, and so at odd integer fillings $\nu= 2n+1$, there is a degeneracy between the states in the $n^{\mbox{th}}$ LL  from the two valleys. These can be split in two ways -- at the single-particle level by application of a valley Zeeman field $\Delta_v$ (e.g., induced by strain), or by interactions between electrons. Accordingly, we work with the Hamiltonian projected to the $n^{\mbox{th}}$ LL, 
  \begin{eqnarray}\label{eq:HFham}
\!\!\!\!\!\!\!\!\!\!\!\!\!\!\!\!\!\!\!\!\!\!\!\!  H &=& \frac{1}{2} \sum_{\kappa,\kappa'} \sum_{{X,Y,X',Y'}} V^{\kappa Y, \kappa' Y'}_{\kappa' X', \kappa X} c^\dagger_{\kappa Y}c^\dagger_{\kappa' Y'}c_{\kappa' X'}c_{\kappa X} +\Delta_v\sum_X \left(c^\dagger_{1X} c_{1X} -c^\dagger_{2X} c_{2X} \right) + E_{{b}}.\end{eqnarray}
Eq. (\ref{eq:HFham}) includes both  a uniform single-particle valley Zeeman term and the Coulomb interaction 
$V^{\kappa X, \kappa' X'}_{\kappa' Y', \kappa Y} = \int d^2r d^2r'\, \psi^{(n)*}_{\kappa X}(\br)\psi^{(n)*}_{\kappa' X'}(\br\,') V(\br-\br\,')\psi^{(n)}_{\kappa' Y'}(\br\,')\psi^{(n)}_{\kappa Y}(\br)$
 where $V(\br) = e^2/\epsilon r$ with $\epsilon$ the 2DEG dielectric constant. $E_b$ denotes interactions between the positive ionic background and electrons as well as the ionic self-interactions. We have omitted two-body ``Umklapp'' terms that cause net  electron transfer between valleys
but are exponentially small in $1/K_0\ell_B$, and inter-valley exchange terms that are suppressed by a factor of $(1/K_0\ell_B)^2$. 

The Hamiltonian (\ref{eq:HFham}) has  $Z_2\times U(1)$ valley-rotation invariance, as well as $U(1)$ charge conservation symmetry. The $Z_2$ corresponds to a $\pi/2$ spatial rotation combined with valley interchange, and is an exact microscopic symmetry of the problem. The $U(1)$ portion of the valley symmetry on the other hand is approximate and stems from neglecting exponentially small intervalley scattering matrix elements in  (\ref{eq:HFham}): it reflects the conservation of valley polarization in the absence of such terms. Note that the two $U(1)$ symmetries may be also viewed as the independent conservation of the valley occupations; it is clear that the intervalley scattering from disorder and/or interactions will conserve only the sum of the two valley occupations, corresponding to the total charge, but not their difference, which corresponds to the $z$-component of valley pseudospin.
As we have argued, in order to form a $\nu=1$ QH state, the ground state must spontaneously break the combined valley symmetry. At this level of analysis, however, we cannot immediately determine whether the system breaks the $Z_2$ or the $U(1)$ symmetry, which must be determined by an explicit energetic computation~\cite{abanin_nematic_2010}. To this end, we variationally optimize Eq. (\ref{eq:HFham}) (taking $\Delta_v=0$ for the moment) using the fully pseudospin polarized trial wavefunction $\ket{\vec{n}} = \prod_X(\cos\frac\theta2 c^\dagger_{1, X} + e^{i\varphi/2} \sin \frac\theta2 c^\dagger_{2, X})\ket{\tilde{0}}$, where $\vec{n} = (\sin\theta\cos\varphi, \sin\theta\cos\varphi, \cos\theta)$ parametrizes a unit vector on the Bloch sphere, and $\ket{\tilde{0}}$ denotes the state obtained by filling both valleys in LLs $0, 1, \ldots n-1$.  Using the Hartree-Fock decoupling of the interactions, it is straightforward to show that the energy per electron is $E[\vec{n}] \equiv \frac{1}{N} \bra{\vec{n}} H \ket{\vec{n}} = - [E_s +  E_a n_z^2]$, where $E_{s,a}= \frac{1}{4N}\sum_{X,X'}  \left(V^{\kappa X, \kappa X'}_{\kappa X, \kappa X'} \pm V^{\kappa X, \bar{\kappa} X'}_{\bar{\kappa} X, {\kappa} X'}   \right)$ are the valley-symmetric and valley-antisymmetric exchange energies, and we have used the valley-interchange symmetry and denoted $2= \bar{1}, 1=\bar{2}$. Note that for a spatially uniform trial state there is no Hartree contribution as this is exactly canceled by the uniform background charge. Explicitly, for the example considered here and for $n=0$, as $N\rightarrow\infty$ we can convert the sums over guiding-center coordinates $X,X'$ to integrals and obtain $E_{s,a} = \Delta_0\frac{C_1\pm C_2}{2}$ where $\Delta_0 = \frac{1}{2} \sqrt{\frac\pi2} \frac{e^2}{\epsilon\ell_B},$ $C_1 = \frac{2}{\pi} \frac{K({1-\frac{1}{\lambda^2}})}{\sqrt{\lambda}}, C_2 = \sqrt{\frac{2\lambda}{1+\lambda^2}}$, and $K$ is the complete elliptic integral of the first kind. Similar computations can be performed in  higher LLs and for other systems given single-particle wavefunctions of the form (\ref{eq:LLwav}). 

Evidently, $E[\vec{n}]$ is azimuthally symmetric on the Bloch sphere and minimized for $n_z = \pm 1$. Therefore, among fully pseudospin-polarized trial wavefunctions, `Ising' states where all the electrons are in one of the two valleys have lower energy than those where the electrons are in a coherent superposition of the valleys. Partially-polarized states at  $\nu = 2n+1$ have a higher energy; this can be attributed to a reduction in the exchange energy gain. 
Hence, the ground state  at $\nu = 2n+1$ is an Ising-ordered QHFM that breaks the $C_4$ rotation symmetry down to $C_2$. Intuitively, in a pseudospin-polarized state, the Pauli exclusion principle forces the wavefunction to vanish when a pair of electrons approach each other, thereby lowering their Coulomb interaction energy; placing all electrons in the same valley does not increase the kinetic energy as it is quenched within a LL. The alert reader will recognize that this argument, which relies only on the fact that the two-body interaction is positive at vanishing separation, is very similar to Hund's rule; here, the quenching of the kinetic energy means that one can view the $N_\Phi$ states within a single LL as degenerate orbitals of a giant `atom'. A more solid-state perspective is that QHFM is analogous to Stoner ferromagnetism, where now the density-of-states singularity stems from the formation of Landau levels. The Ising anisotropy results from the microscopic interactions  that distinguish between the two valleys, in turn a consequence of the spatially anisotropic dispersion. It vanishes when the valley dispersions become isotropic ($\lambda\rightarrow 1$) or if the two valleys have the {\it same} anisotropy and hence $V^{\kappa X, \kappa X'}_{\kappa X, \kappa X'} =  V^{\kappa X, \bar\kappa X'}_{\bar\kappa X, \kappa X'}$; in such cases the QHFM has $SU(2)$ symmetry as does the effective Hamiltonian (\ref{eq:HFham}). We are unaware of any examples of a valley QHFM with $U(1)$ symmetry; for this to occur, short-range interaction corrections must cause the  intervalley  exchange  to dominate the intravalley contribution, which seems physically unlikely. Indeed, for the case of pure Coulomb interactions there is a general proof that the fully-valley-polarized solutions are the lowest-energy trial states~\cite{Sodemann2017}.
 Thus  this simplest instance of a valley QHFM has Ising-nematic order: {\it Ising} describes the easy-axis symmetry of the energy functional in order-parameter space, while {\it nematic} reflects the orientational symmetry breaking.

In the presence of a spatially anistropic but uniform external potential, such as that from uniaxial strain imposed using a piezoelectric sample stage~\cite{ShayeganStrain}, there will be an effective splitting between the two valleys already at the single-particle level and for $B=0$. 
When projected onto a LL for $B\neq 0$, this gives rise to a valley Zeeman splitting $\Delta_v$ as in the second term in (\ref{eq:HFham}). Observe that for our trial state $\ket{\vec{n}}$ this simply leads to a contribution $\propto \Delta_v n_z$: spatial anisotropies couple directly to the ($z$-component) of the QHFM order parameter, reflecting its nematic nature.  Finally, quenched randomness can also be incorporated microscopically by adding  to (\ref{eq:HFham}) a term of the form $H_{\mbox{r}} = \mathcal{P}_{n} \left(H_{st} +  H_{pot}\right)\mathcal{P}_n$, where $H_{st}, H_{pot}$ denote contributions of random strains and random potentials, respectively,  and $\mathcal{P}_{n}$ denotes projection to the $n^{\mbox{th}}$ LL. While analyzing the random strain is straightforward and leads to a spatially-dependent valley Zeeman field, the random potential is more subtle. Crucially, local anisotropies in the random potential couple as a random valley Zeeman contribution; extracting this requires  studying virtual processes that involve higher Landau levels. This involves a technical but not particularly illuminating computation~\cite{KumarSAPSLSDWs}, so we omit it here and simply summarize its consequences for the effective field theory in the next section.

The above microscopic approach can be used to compute the parameters of the low-energy effective theory that describes deviations from a uniform QHFM state. The calculation proceeds (for an explicit example for the bilayer case, see~\cite{Moon:1995p1}) by considering textured states that are obtained by restricting valley pseudospin-wave perturbations about a uniform state to the $n^{\mbox{th}}$ LL. 
In other words, we perform a type of  `single-mode-approximation' by considering a spatially varying trial state of the form $\ket{\vec{n}_{\bq}} = e^{-i\sum_{\bq} \vec{n}_{-\bq} \cdot\overline{\vec{S}_{\bq}}}\ket{\vec{0x}}$. Here 
$\overline{S^\mu_{\bq}} = \sum_{X\kappa\kappa'} e^{iq_x X}  c^\dagger_{\kappa,X_+ }[\frac{\sigma^\mu}{2}]_{\kappa\kappa'} c^{\phantom\dagger}_{\kappa'X_-}$
with $X_{\pm} = X\pm \frac{q_y\ell_B^2}{2}$ represents the pseudospin operator projected the LL of interest. In essence, acting with $\overline{S^\mu_{\bq}}$ on a uniform ground state produces a `twist' in the local pseudospin with amplitude $n_{\bq}$ at wavevector $\bq$, but within the restricted subspace of the LL. Recall that the Hamiltonian (\ref{eq:HFham}) may be written in terms of projected density operators\footnote{Technically the operators $\rho_{\bq}$ are actually magnetic translation operators that differ from the true projected density by a LL form factor, but for present purposes this distinction is unimportant.}  
$\overline{\rho_{\bq}}= \sum_{X\kappa} e^{iq_x X}  c^\dagger_{\kappa,X_+ } c^{\phantom\dagger}_{\kappa X_-}$; the key physical point is that charge and pseudospin are coupled when projected to the LL. Thus, modulating the spin in the LL induces a charge density, which leads to a Coulomb interaction energy. As long as we are interested in long wavelength properties, we may compute the energy of the modulated trial state order-by-order in $\bq$ --- i.e., via a gradient expansion.  In doing this, it is especially convenient to leverage the  (generalized) Girvin-Macdonald-Platzmann algebra~\cite{GMP:1986p1,Moon:1995p1} of the projected charge and spin operators,
 \begin{eqnarray}\label{eq:GMP}
 [\overline{\rho_{\bq_1}}, \overline{\rho_{\bq_2}}] = 2i \sin\frac{\bq_1\wedge\bq_2}{2} \overline{\rho}_{\bq_1+\bq_2},  [\overline{\rho_{\bq_1}}, \overline{S^\mu_{\bq_2}}] = 2i \sin\frac{\bq_1\wedge\bq_2}{2} \overline{S^\mu}_{\bq_1+\bq_2}, \nonumber\\
\left[\overline{S^\mu_{\bq_1}}, \overline{S^\mu_{\bq_2}}\right] = \frac{i}{2}\delta^{\mu\nu} \sin\frac{\bq_1\wedge\bq_2}{2} \overline{\rho}_{\bq_1+\bq_2} + i \epsilon^{\mu\nu\sigma}\cos\frac{\bq_1\wedge\bq_2}{2} \overline{S^\sigma}_{\bq_1+\bq_2}.
 \end{eqnarray} 
However, if we are uninterested in computing the values of the coefficients from first principles, we can simply infer the low-energy effective theory on symmetry grounds, a task to which we turn in the next section. We will return to the microscopic approach when considering properties of domain walls in Sec.~\ref{sec:topodeftransport}. 
 
\subsection{Relevance to Other Systems}
So far, we have specialized to the two-valley case relevant to AlAs. The minimal theory developed in this section and the next remains relevant to more complex multivalley 2DEGs~\cite{Ando} in some situations; we briefly sketch these here, as this allows us to streamline the discussion of present experiments, which can largely be explicated with the tools at hand, while deferring a full discussion of the new physics that can arise with more valleys to Sec.~\ref{sec:complexity}.

For specificity, we will consider systems with $\mathcal{N}$ valleys that are identical in shape but have their symmetry axes oriented by $2\pi/\mathcal{N}$ with respect to each other. The band structures of several such systems are shown schematically in Fig. \ref{fig:ModelBandStructures}. We will also ignore sets of $\mathcal{N}$ filled LLs, so that all fillings are assumed to be modulo $\mathcal{N}$ (Of course for high enough LLs and in systems with complicated dispersion, other possibilities may emerge due to microscopic details of the Hamiltonians). With these caveats, the applicability of the minimal model may be summarized as follows:
\begin{figure}
\begin{centering}
\includegraphics[width=.75\columnwidth]{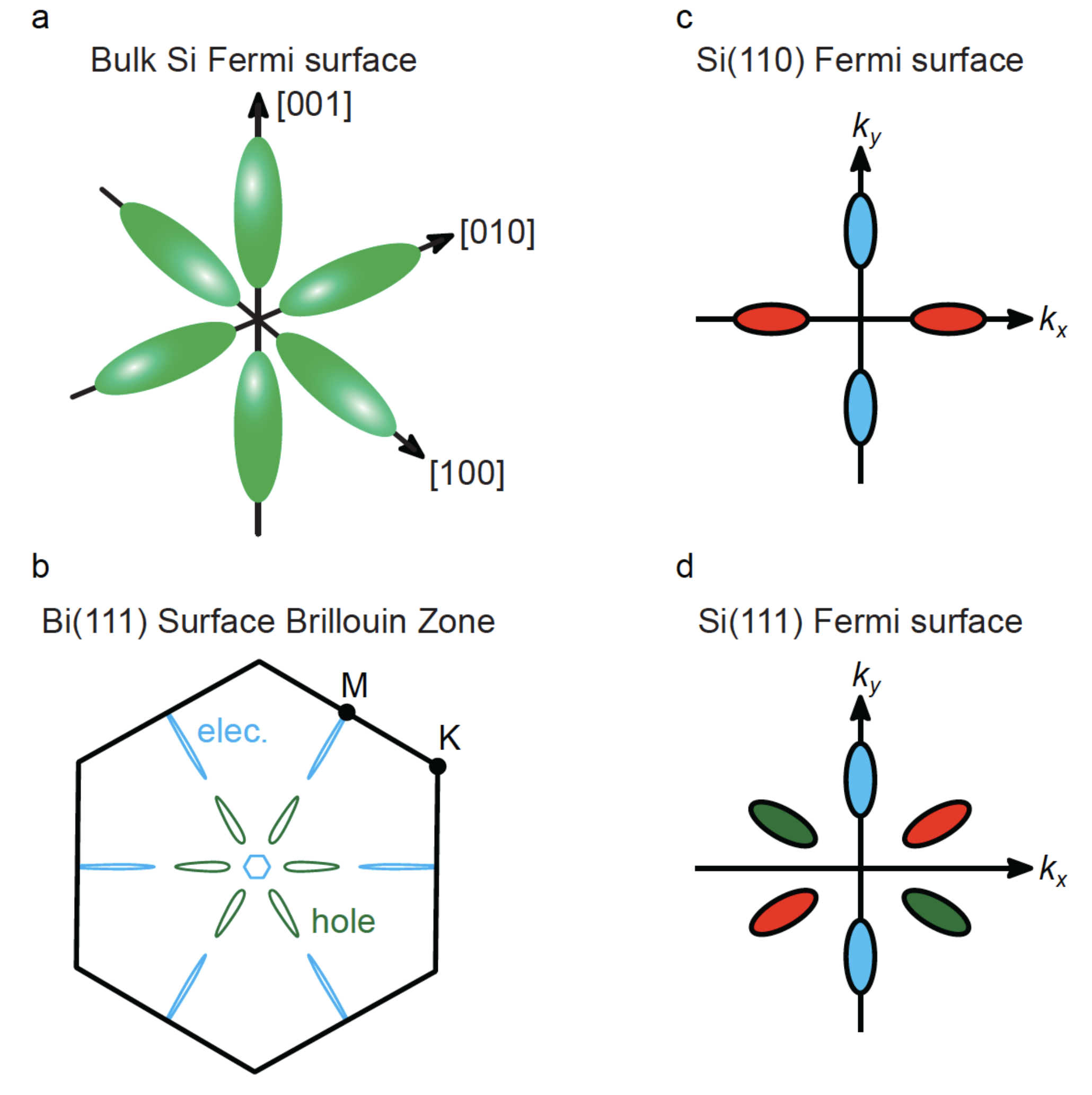}
\end{centering}
\caption{Some model band structures discussed in this review. \label{fig:ModelBandStructures}}
\end{figure}
\begin{enumerate}
\item Our work immediately generalizes, with suitable modifications, to the $\mathcal{N} =3$ case for $\nu=1,2$. The resulting effective theory will be more complex now as the order parameter is a ${Z}_3$ variable. The role of  strain is also more subtle, as it can split the threefold degeneracy either completely or leave a residual twofold degeneracy, depending on the axis along which it is applied. Note that in crystals that have only threefold rotational symmetry, the three valleys are entirely contained within the BZ.  In systems with sixfold rotational symmetry, the three valleys must be centered at the $M$ points of the hexagonal Brillouin zone, as is the case for the (111) surface of SnTe~\cite{LiZhangMacDonald}.
\item In the  $\mathcal{N} =4$ case all the valleys must be contained within the BZ. In this case, it is easy to see that perfectly elliptical valleys rotated by $\pi$ with respect to each other have identical anisotropy; we dub these ``anistropy partners". Following the energetic computations above we would conclude that QHFM states obtained by rotating between anistropy partners are related by SU(2) symmetry, complicating the situation. For $\nu=1$, QHFMs will not only break orientational symmetry but {\it also} this SU(2) symmetry, resulting in a ``valley wave''~\cite{Rasolt:1986p1},  Goldstone mode, that can only be captured by a more complex theory\footnote{In the microscopic case, via an RPA calculation about the Hartree-Fock trial state; in the effective field theory, by a more intricate sigma model.}. At $\nu=2$ an explicit calculation reveals that the leading interaction terms do not split the degeneracy between filling both LLs in an anisotropy pair or filling one LL from each pair, and these possibilities break distinct symmetries. While in very clean systems with no external strain resolving this degeneracy can be a subtle issue [see Sec.~\ref{sec:complexity}], in many cases uniaxial strain will favor the first possibility. Then since the resulting QHFM only breaks a discrete symmetry (the $SU(2)$ symmetry is no longer broken when both states in a pair are filled), our minimal $Z_2$ model suffices to capture the essential physics. The results for $\nu=3$ can be argued to be identical to those at $\nu=1$ by particle-hole symmetry within the LL.  This scenario is relevant to Si(110) quantum wells~\cite{Ando}.
\item For  $\mathcal{N} =6$ again all the valleys must be contained within the BZ, and now there are three pairs of $SU(2)$-related LLs. For $\nu=1$ a QHFM  state again breaks orientational symmetry, but also $SU(2)$, and hence requires a more complex treatment than considered thus far. For $\nu=2$, we also find an exact degeneracy  as in the preceding case; assuming again that strain lifts the degeneracy between different anisotropy pairs, we can then simply rewrite the problem in terms of filling distinct anisotropy pairs, reducing it to the $Z_3$ case.  For $\nu=3$ there is yet again a residual degeneracy between different classes of states, but even if strain lifts the degeneracy, one anistropy pair will always have only one LL's worth of electrons. Therefore, the final QHFM always has to break the SU(2) symmetry of rotations within this pair, leading to a Goldstone mode not captured by the minimal model. Results for $\nu =4,5$ follow by particle-hole conjugation of the $\nu=2,1$ cases. This scenario applies to Si(111) quantum wells~\cite{Ando,eng_integer_2007,mcfarland_temperature-dependent_2009,kott_valley-degenerate_2014}
and to surface states of bismuth on its (111) surface~\cite{Feldman2016,Randeria2018}. Note that in each case additional corrections (e.g., `teardrop' anisotropies in each valley) and spin texture of pairs of valleys may lower the symmetry between valleys in an anisotropy pair from the full SU(2) case (see Sec.~\ref{sec:complexity}.). This distinction is generally unimportant at even $\nu$.
\end{enumerate}
In cases with, e.g., warping there can be additional valleys that emerge to flank high-symmetry points in the BZ; these will typically require separate consideration, although many features may be shared with the above examples.  This situation is relevant, for instance, to bilayer (or higher multilayer) graphene in the presence of trigonal warping~\cite{PhysRevLett.96.086805,PhysRevB.80.165409}
Recent experiments have begun probing the associated physics in trilayer graphene~\cite{Zibrov2018}. Also, specific realizations can display further modifications of the interaction form factors, e.g. due to spin-orbit coupling effects. These and other symmetry considerations can also have substantial consequences  for understanding how external fields couple to the order parameter and/or break some of the valley symmetries (this is beyond the scope of this review, but for a good example, see~\cite{LiZhangMacDonald}). Observe that the SU(2) symmetry in QHFMs that spontaneously break symmetry between two anisotropy partners is typically an added level of complexity {\it beyond} nematic order. The fillings where our minimal model applies are precisely those that isolate nematicity as the sole symmetry-breaking mechanism, and we will focus on this setting for the next two sections, returning to the more involved examples in Sec.~\ref{sec:complexity}.

\section{Effective Field Theory \label{sec:EFT}} 

While the microscopic approach is necessary in order to  determine the anisotropy of the QHFM and in computing precise energy scales, for an intuitive understanding of the problem it is convenient to work with an effective field theory. In common with other examples of QHFMs, in the minimal $\mathcal{N}=2$ case this takes the form of a nonlinear sigma model (NLSM) for the nematic order parameter $\vec{n}$. There are two key symmetry principles that govern the effective action, namely (i) the Ising symmetry in order parameter space and (ii) the coupling of $n_z$ to spatial anisotropy. The underlying exchange physics favors spatially uniform order parameter configurations, captured by a nonzero stiffness $\rho_s$.   At leading order in a gradient expansion, these criteria lead us to the classical energy functional 
\be\label{eq:classen}
E[\vec{n}({\br})] =  \int d^2r\,\left[ \frac{\rho_s}{2} (\nabla \vec{n})^2  - \frac{\alpha}{2}n_z^2 + (h_0 +  h_r({\br})) n_z({\br})\right],
\ee
where, for the $n=0$ case and in our two-valley example,  $\rho_s = \frac{e^2}{16\sqrt{2\pi} \epsilon \ell_B} + O(\lambda-1)$ and $\alpha \approx \frac{3}{32} \frac{\Delta_0}{2\pi \ell_B^2}(\lambda-1)^2$. These may be derived microscopically as described in the preceding section. Note that in this section we will assume that $|\lambda-1| \ll 1$; this is a theoretical convenience as it enables us to focus on the essential physics~\footnote{For instance, we may ignore spatial anisotropies in the stiffness as these are negligible in this limit.}. Even though experimental systems have strong anisotropy, the picture from the weak-anisotropy limit is qualitatively similar, and is often very useful for making intuitive arguments~\cite{abanin_nematic_2010}. 

There are two contributions to the conjugate field $h$ that couples to the order parameter. A spatially uniform piece $h_0$ is assumed to be due to a uniaxial external strain  while the random piece $h_r({\br}) =h_{st}({\br}) + h_{pot}({\br})$  receives contributions  both due to random strain $u({\br})$  and due to anistropies in the smooth  random potential $U(\br)$ generated  by disorder in the 2DEG (cf. $H_{st}, H_{pot}$ from the preceding section). Explicitly, we have  $h_{st}({\br})  \propto \frac{\partial u}{\partial x}-\frac{\partial u}{\partial y}$, where $u(\br)$ is the displacement of point $\br$ in the crystal, and  $h_{pot}({\br}) = \frac{(m_x -m_y)\ell_B^2}{2\pi\hbar^2} \left[(\partial_xU)^2 - (\partial_y U)^2\right]$; the latter can be justified via a microscopic calculation that incorporates Landau-level mixing. (This also gives a secondary contribution that vanishes except at domain boundaries and has therefore been omitted.) In the following, we will  assume that $h_r$ has zero mean and is short-range correlated, with $\int d^2 r\langle h_r({\br})h_r({\mathbf{0}})\rangle = W$.

In order to obtain a full quantum description, (\ref{eq:classen}) must be augmented with a kinetic term, which takes the form of the usual Berry-phase contribution for a ferromagnet. In addition, a central aspect of QHFMs is that spin textures with topological charge  --- such as skyrmions --- also carry electrical charge. A microscopic justification for this can be given using the methods of the preceding section, and is reflected in the nonvanishing density-spin commutator in (\ref{eq:GMP}). Therefore in order to properly treat the presence of such charged spin textures, the energy functional must also include Coulomb interactions between topological charges located at different positions, as well as a topological (Hopf) term to encode the fermionic statistics of the skyrmion. However, as we shall argue below, in the Ising case such topological charges play a limited role and so we omit them from the outset for simplicity.

Since we are in two dimensions and the Ising-nematic state breaks a discrete symmetry, this occurs at a finite transition temperature $T_c$ in the absence of disorder. The energy functional  (\ref{eq:classen}) allows us to make a simple estimate of the nematic ordering temperature; observe that (\ref{eq:classen}) is the continuum limit of a 2D Heisenberg ferromagnet with weak Ising anistropy, whence we find that $k_BT_c \sim 4\pi \rho_s/\log[\rho_s/\alpha\ell_B^2]$. Intuitively, this follows from the fact that $T_c$ should vanish as $\alpha\rightarrow 0$ (since the symmetry in that limit is continuous and hence precluded from breaking at finite temperature by the Mermin-Wagner theorem), though this vanishing occurs only logarithmically slowly since $d=2$ is the lower critical dimension for the classical Heisenberg model. This estimate is predicated on a weak-anistropy limit, but since in reality the anisotropy is not small, it is reasonable to simply set $T_c\sim \rho_s$, which is similar to that obtained by a microscopic Hartree-Fock calculation. For the $n=0$ LL of AlAs, and in field ranges from 1 T to 10 T characteristic of most QH experiments, $T_c$ is typically in the range of several Kelvin. As this is well above the typical temperatures (ranging from $\sim$ 10-100 mK) at which most QH experiments are carried out, it may be feasible to see features of this transition in experiments. Quenched disorder can complicate this, since it is both ubiquitious and a relevant perturbation (in the renormalization-group sense) to the Ising-ordered phase in $d=2$. We turn now to an analysis of its effects.

\section{Quenched Disorder, Domain Formation, and Single-Domain Physics\label{sec:disorder}}
In the previous section, we argued for the existence of a finite-temperature Ising transition into the valley-nematic QHFM phase. This ignored the role of quenched disorder; its inclusion can significantly alter the story. To understand this, it is convenient to leverage universality and replace our continuum field theory with a classical $d$-dimensional lattice Ising ferromagnet in a random field, $H = -J\sum_{\langle ij\rangle}s_i s_j - \sum_i h_i s_i$, where the $s_i = \pm1$ are classical Ising spins arranged on a regular lattice, and the $h_i$ are random fields that are drawn from a distribution of zero mean and standard deviation $W$. When $W=0$, this model has a finite-temperature transition at $T_c\sim J$, and for $T\ll T_c$, the dominant configurations of the system are those where all the spins are aligned. A classic argument~\cite{imry_random-field_1975} considers the fate of the $T\rightarrow 0$ ordered phase when $W\neq 0$, and proceeds as follows. A sufficiently strong local Ising field may favor the formation of a local field-aligned domain, despite the fact that this involves a surface energy cost due to misaligned spins at the boundary.
The energy gain for creating a domain of linear dimension  $L$ is controlled by the average field over the domain, $E_h \sim  - c_1 WL^{d/2}$, whereas the energy cost is governed by the Ising energy, $E_{J}(L) \sim c_2JL^{d-1}$ (Here $c_{1,2}$ are nonuniversal constants of order unity). It is therefore evident that while the ordered phase is stable against domain nucleation for $d>2$, it is destroyed for arbitrarily small $W$ for $d<2$. In the $d=2$ case of interest to us both terms are of the same order in $L$ so this simple argument is inconclusive. A more general solution~\cite{binder_random-field_1983,Aizenman:1989p1} wherein the domain walls are allowed to `roughen' in response to the local field admits a length scale $\xi_B \sim e^{-c J^2/W^2}$ (where $c$ is a  constant of order unity), beyond which the field energy dominates. Since there is always some disorder present in any sample, we conclude that as a matter of {\it principle} the Ising-nematic phase is destroyed in the thermodynamic limit.

Despite this point of principle, since typical domain size $\xi_B$ is exponentially sensitive to disorder, in {\it practice} this destruction of the ordered phase can be of limited relevance to experiments. The situation is similar to that of Anderson localization in 2D: while all states are indeed localized in the thermodynamic limit, the exponential sensitivity of the localization length means that many even modestly disordered experimental samples remain metallic as they are short compared to the localization length. In our case, the disorder can be extremely dependent both on the sample preparation, as well as on the specific material setting; therefore, it is not unreasonable to anticipate that $\xi_B$ can vary by an order of magnitude or more in a given system, and exhibit much bigger variations between different materials. Samples with typical size $L_s \gg \xi_B$ will indeed show multi-domain structure and will be in the disorder-dominated limit; samples of size $L_s\ll \xi_B$ will typically contain a single domain. Although we will focus on these two limiting scenarios, we note that there is also an interesting `mesoscale' regime of samples with a few large domains, that could present additional interesting features for experiment.

\subsection{Transport Probes of Domain Formation}
A natural question  is whether it is possible to detect this domain structure experimentally. In this setting, it is important to disentangle the role of two distinct and typically independent sources of disorder that are present in most samples. The first is the smooth disorder potential, such as that induced by donor ions in a dopant plane, which is typically offset a finite distance $d \gg \ell_B$ from the 2DEG; this combines with the local random strains to produce the effective random field discussed in the preceding section. In addition, there may also be atomic scale defects in the immediate vicinity of the 2DEG plane, whose potential is sharp relative to the scale of the magnetic length. These will typically have a characteristic scattering length $a_0\ll \ell_B$ and may be approximated as $\delta$-function potentials.  
At the single-particle level sufficient to capture the relevant physics, the impurity potential shifts the energy of only a single state localized near the impurity, leaving the other states in the LL untouched. This results in the formation of a band of localized states from each of the valleys. Crucially, in the QHFM setting these shifts can
be small relative to the exchange splitting between different valley-polarized states, resulting in the formation of a narrow band of impurity states all with the same anisotropy, split by 
$\sim e^2/\epsilon\ell_B$ from states from the opposite valley.
Of course, within each impurity band there will be at least one extended state, as required by topological considerations for nonzero $\sigma_{xy}$~\cite{PhysRevB.25.2185}, but this is unimportant to our discussion below.

In single-domain samples, $\sigma_{xy}$ is quantized, and at small but nonzero temperature there is also a {longitudinal} resistivity  due to hopping between these impurity bound states. The latter has  activated or stretched-exponential scaling with temperature, depending on whether nearest-neighbor or variable-range hopping dominates. At this point we must  clarify the structure of the individual low-energy charged quasiparticle excitations. In the weak-anisotropy limit $\lambda\rightarrow 1$,  $\alpha\ll \rho_s\ell_B^2$. In the SU(2) limit, it is known the lowest-energy charged bulk quasiparticle is a Skyrmion, that acquires a finite size due to corrections from Zeeman effects~\cite{SondhiSkyrmion}. Ref.~\cite{Sodemann2017} suggested that such a scenario may also apply for the case of small Ising anisotropy for pure Coulomb interactins. The precise determination of the anisotropy at which skyrmions become unfeasible in a particular system lies in details of the short-range interactions that are beyond the scope of the present discussion. However, such a `valley skyrmion' is likely to have a very small valley polarization (in contrast to the case of spin skyrmions in GaAs, whose spin can be substantial~\cite{PhysRevLett.75.4290,PhysRevLett.74.5112,Tycko1460}), and would therefore appear quite similar to single pseudospin-flip charged excitations, at least at the level of their activation gap and transport properties.
In the balance of this review, unless otherwise specified we will understand the term ``charged quasiparticle''  to refer to the single pseudospin-flip excitation, and explicitly refer to ``valley Skyrmions'' when necessary to discuss those.

Returning to our hopping problem, a  crucial fact is that quasiparticles in the two valleys have distinct contributions to the resistive anisotropy. For hopping exclusively between quasiparticles in valley $\kappa$, a  scaling argument~\cite{PhysRevB.48.11492} reveals that the resistive and mass anisotropies are simply related $\sigma_{xx}/\sigma_{yy} \propto {1}/{u^2_{\kappa}}$.  The dimensionless ratio
\be
N =  \frac{\sigma_{xx} -\sigma_{yy}}{\sigma_{xx} +\sigma_{yy}} 
\ee
is a measure of this anisotropy, with $N_\kappa = \frac{1-u_\kappa^2}{1+u_\kappa^2} = -N_{\bar{\kappa}}$ for hopping between states in the two valleys.
When the chemical potential is in the middle of the $\nu=1$ Hall plateau, localized states from both valleys are typically present, with identical density of states --- the latter a consequence of the combined particle-hole/valley-reversal symmetry of the state in the absence of LL mixing~\footnote{Note that realistic disorder will typically also break particle-hole symmetry, and therefore modify this picture; despite this we expect qualitatively similar features to remain even in this case.}. When $\nu\neq 1$, particle-hole symmetry is broken. Let us assume that the filled states are in valley $\kappa$. Then, at filling factor $\nu = 1-\delta\nu$ with $1\gg \delta\nu \gg 0$,  the density of localized states in valley $\kappa$ exceeds the density of states in valley $\bar{\kappa}$; since hopping resistivity is exponentially sensitive to the density of states, we expect that the valley-$\kappa$ hopping dominates, leading to $N \approx N_\kappa$. Running a similar argument for $\nu = 1+\delta\nu$ we find that now $N\approx N_{\bar\kappa} = -N_{\kappa}$. Thus, $N(\nu)$
changes sign as $\nu$ is tuned across the center of a $\nu=1$ nematic QHFM plateau in a single-domain sample. In experiments where uniaxial strain can be externally imposed~\cite{ShayeganStrain}, 
it should also be possible to tune a sample into a monodomain regime, where the anisotropy $\sigma_{xx} \neq \sigma_{yy}$ is especially prominent~\cite{Shayegan2006}.
Note that we have implicitly assumed a small but non-zero temperature in this discussion, since at $T=0$ the longitudinal conductivity vanishes and therefore $N$ is ill-defined. (Since $\sigma_{xx}$ and $\sigma_{yy}$ share the same temperature dependence, $N$ itself should have a finite zero-temperature limit.)  Alternatively it may also be possible to perform measurements at small but finite  frequency $\omega$ and then take the $\omega\rightarrow 0$ dc limit. It should be clear that since $\lambda^2$ is typically at least $O(1)$, the extremal values of $|N|\sim \frac{\lambda^2-1}{\lambda^2+1}$ can be appreciable. A similar argument applies at any ferromagnetic filling factor $\nu=p$ such that the LLs immediately below and above the chemical potential have distinct anisotropies.

\subsection{Impurities as Local Probes of Nematic Order}
The role of atomic scale impurities is also evident in more recent STM measurements of nematic states on the Bi(111) surface~\cite{Feldman2016}. Local spectroscopic measurements of this material show that the six-fold valley degeneracy of the hole surface states is lifted by a combination of strain and exchange interactions to produce three valley-polarized states at even integer fillings.  Each of these even-integer-filling states corresponds to filling both valleys that share the same spatial orientational anisotropy; this choice is imposed on the system by the extrinsic symmetry-breaking, but absent such contributions other ordered states are possible (see Sec.~\ref{sec:complexity} below). An STM tip may be used to spatially map the differential conductance, which is in turn proportional to the local density of states (LDOS). Experiments reveal concentric elliptical features with a different preferred directionality for each broken-symmetry LL peak. These low-conductance ellipses are centered around the atomic-scale impurity sites,   and hence the data can be interpreted as the imaging of individual isolated LL states 
 that are pinned to the impurities.  The anisotropy of each hole valley on the bismuth surface results in a corresponding anisotropic cyclotron orbit. Therefore, STM imaging allows for a direct visualization of the preferred wavefunction directionality and hence of the local nematic order. Long-wavelength disorder appears less prominent and there are quite large nematic domains; at present, the role of ``terrace'' effects due to exfoliation in pinning domains remains unclear.

\section{Topological Defects and Transport in Multidomain Samples \label{sec:topodeftransport}}
 So far, we have discussed the physics of single-domain samples that show the characteristic properties of QH Ising nematic order. As we have argued, smooth disorder will lead to domain formation, and for sufficiently large samples or sufficiently strong disorder, the ordered phase will break up into many domains. In this regime, the nematic response will be averaged out and so the local valley Ising order transport will be invisible to transport; nevertheless, spatially resolved probes such as STM can directly visualize the nematic domains. In multidomain samples, it is essential to consider the role of topological defects of the QHFM, to which we now turn.

Topological defects in QHFMs have a rich history. Owing to the  quantized Hall conductance they are endowed with a set of unusual features --- for instance, skyrmion textures carry an electrical charge, and are the lowest-energy charged excitations in spin QHFMs in the limit of weak Zeeman coupling. 
In Ising QHFMs, the relevant 
topological defects are domain walls between different Ising domains. In the present nematic setting, each of these domains will break orientational symmetry in a distinct fashion --- for example, in the $\mathcal{N}=2$ case relevant to AlAs the two distinct domains are those where the QHFM has electrons polarized in  valley $1$ and valley $2$, respectively.

\begin{figure}
\includegraphics[width=\columnwidth]{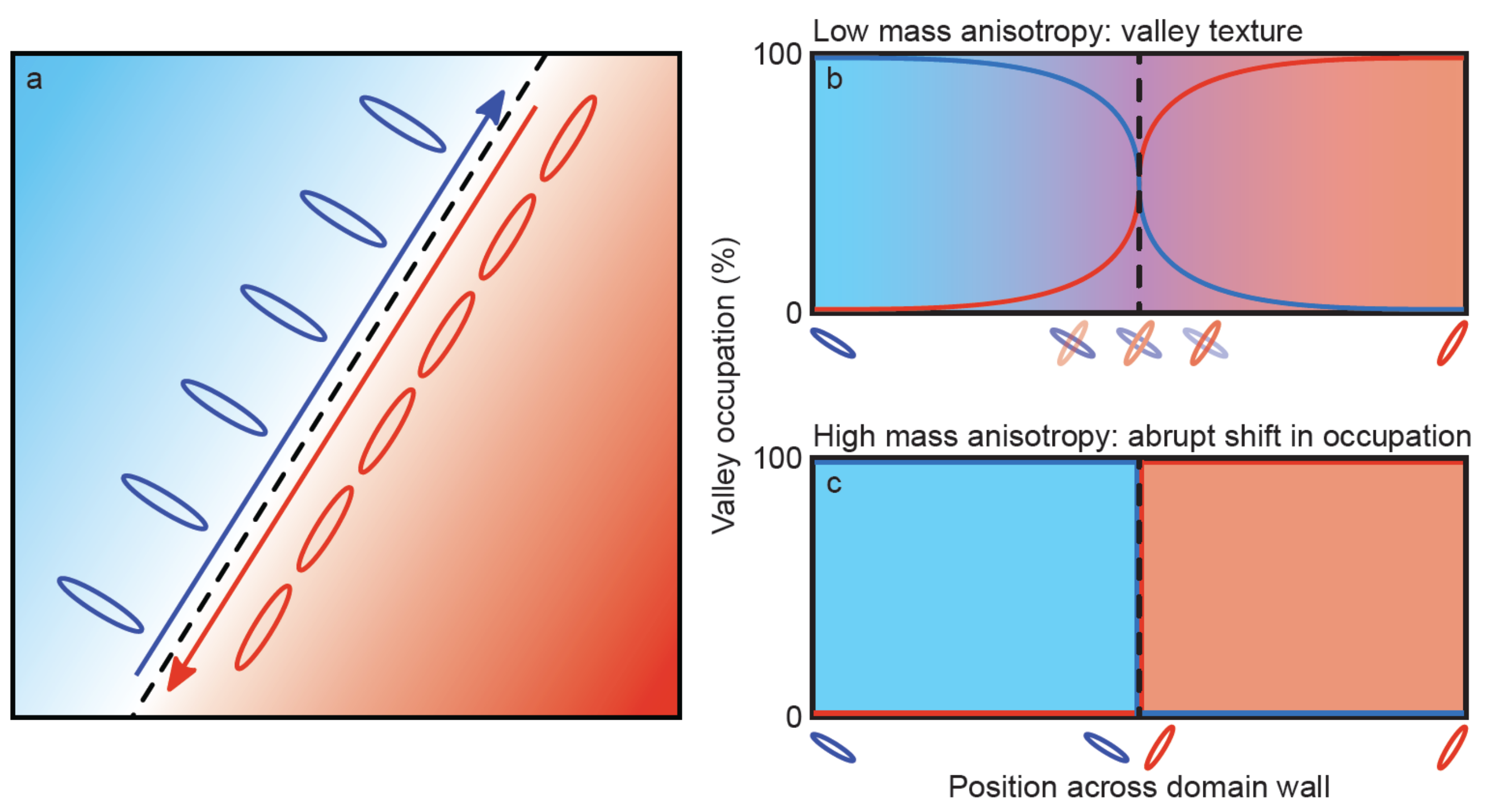}
\caption{Nematic Domain Walls. (a) Domain walls form between regions of distinct  polarization of the nematic order parameter and host counter-propagating, valley-filtered one-dimensional modes. (b) When valley anisotropy is small, the domain wall can show `texturing', so that in an appreciable region near the domain wall there is nontrivial inter-valley coherence --- i.e., the valley pseudospin order parameter points along the equator of the Bloch sphere. (c) For large anisotropy, the domain wall is `sharp' --- the valley occupation changes abruptly (on the scale of a magnetic length).
\label{fig:DomainWall}}
\end{figure}

Within the effective field theory, a domain-wall parallel to the $y$-axis is obtained by minimizing the classical energy (\ref{eq:classen}) while setting the boundary condition $n_z = \pm1$ for $x\rightarrow \pm \infty$. The resulting solution has a  domain-wall width $L_0 \sim \sqrt{\rho_s/\alpha}$ that characterizes the region where the order parameter lies in the $xy$ plane  (Fig. \ref{fig:DomainWall}b), and a surface tension (energy per unit length parallel to the wall) $\mathcal{J}\sim \sqrt{\rho_s\alpha}$.  The domain wall solution spontaneously breaks the residual $U(1)$ symmetry since the energy is invariant under rotations in pseudospin space about the $n_z$ axis. Since this putative symmetry breaking occurs along a one-dimensional domain wall, quantum and/or thermal fluctuations will restore the symmetry, but for $T=0$ there is a residual gapless `almost-Goldstone' mode and the XY component of the order parameter shows algebraic correlations along the domain wall. Crucially, phase slips of this order parameter carry a topological charge that is translated into an electrical charge via the quantized $\sigma_{xy}$.  An alternative and more microscopic picture is given by a variational ansatz for the domain wall of the form
\begin{equation}\label{eq:DWtrial}
\ket{\mbox{DW}} = \prod_{X} (u_X c^\dagger_{1,X} +\sqrt{1-u_X^2} \,\, c^\dagger_{2,X})\ket{0}, \mbox{ with } u_X = \left\{ \begin{array}{cc} 1, & X\rightarrow-\infty\\ 0, &X\rightarrow+\infty\end{array} \right..\,\,\,\,\,
\end{equation} 
 In the strong anisotropy limit, we find that $u_X$ jumps discontinuously between its two limiting values across the domain wall, so that the domain wall width is simply set by the magnetic length $\ell_B$.  In this limit, we can view the domain wall as being the edge of two distinct QH liquids: a $\nu=1$ state in valley 1 for $X<0$, and a $\nu=1$ state in valley $2$ for $X>0$. Therefore, it hosts a pair of counterpropagating gapless edge modes distinguished by their valley index, so that in the presence of interactions and the absence of intervalley scattering the effective description of these modes is in terms of a gapless Luttinger liquid, with backscattering interactions forbidden by valley symmetry. (Fig. \ref{fig:DomainWall}a). 
  An effective Luttinger liquid theory for the domain wall modes can also be derived from the sigma-model description, and thus we have two complementary treatments of the domain wall. By explicitly evaluating the ground state energy using (\ref{eq:DWtrial}) while allowing $u_X$ to vary spatially~\cite{KumarSAPSLSDWs}, we can recover the sigma-model description of the domain wall. {\it Inter alia}, this calculation reveals that  the `sharp' domain wall is a more appropriate description for mass anisotropies in the range relevant to AlAs or Bi. (Fig. \ref{fig:DomainWall}c).
  In both cases it is possible to show~\cite{Nu2DWTheory} that the  Luttinger parameters and collective mode velocities are controlled by a combination of the interaction strength, the anisotropy energy,  and the strain gradient that serves to pin the position of the domain wall to a specific spatial location~\cite{Falko:1999p1,Mitra:2003p1}. 

The domain wall excitations apparently provide a means of gapless charge transport; were this indeed the case then their presence would immediately lead to a finite longitudinal conductivity and the destruction of the QHE in multidomain samples. Two different mechanisms could alter this conclusion:
\begin{itemize}
\item In disordered samples with a large impurity density, valley-mixing impurity scattering can produce backscattering between the two counterpropagating modes of the Luttinger liquid, leading to localization. In this case, charge transport via the domain wall channel is obstructed and the algebraic correlations are replaced by an exponential falloff over a length scale $\xi_{iv}$ (an effective localization length). Thus, the longitudinal conductivity due to the domain wall modes vanishes exponentially as $T\rightarrow 0$ owing to the formation of mobility gap in the disordered limit. This disordered QH state was dubbed the quantum Hall random-field paramagnet (QHRFPM)~\cite{abanin_nematic_2010}.
\item The intervalley scattering contributions neglected in (\ref{eq:HFham}) could potentially open a mini-gap even in {\it clean} samples. Here, a careful analysis~\cite{Nu2DWTheory} shows that the allowed terms in the resulting effective description when the QHFM is at $\nu=1$ correspond to purely forward scattering, and therefore do not open a gap. The only interactions capable of gapping the system are exponentially suppressed to $O(e^{-\ell_B/a})$, a consequence of 2D momentum conservation in the 2DEG. However for $\nu=2$ QHFMs (that emerge for $\mathcal{N}>3$)
 domain walls carry two sets of counterpropagating modes that can in certain cases be coupled by `double umklapp' processes that conserve momentum in the 2D BZ. These can then open a charge gap leading to a novel charge-valley separated gapless state along the domain wall, that is a charge insulator but a thermal metal. The precise nature of electrical and thermal transport in systems with multiple domains but negligible impurity scattering at $\nu=1,2$ remains an active subject of study. The emergence and protection of the gapless edge modes are linked intimately to the physics of symmetry-protected topological phases and may be viewed as an example of the phenomenon of `anomaly inflow'~\cite{CALLAN1985427}.
\end{itemize}
If the domain wall is a charge insulator --- whether due to disorder or due to interactions --- multidomain samples also exhibit the QHE in charge transport, though the observable signatures of transport are quite different as compared to the single domain case, as we will discuss shortly.

\subsection{STM Spectroscopy of Nematic Domain Walls}
STM measurements offer a method to directly visualize the presence of nematic domains and, in principle, explore the behavior of the associated domain walls. Different nematic regions, with a typical size of order 1 $\mu$m or larger, have been experimentally identified on the surface of bismuth based on measurements of cyclotron orbit orientations~\cite{Feldman2016}. In different areas of the sample, the data reveal different sequences in energy of valley occupation. This guarantees that LLs associated with different valley anisotropy must cross as the domain wall is traversed. Such behavior has been difficult to isolate because all domain walls reported to date using this technique occur in the vicinity of pronounced strain defects as well as near step edges, which reduce the LL visibility. Thus, while significant variations in the energies of the LLs associated with different valleys were observed, full characterization of the expected boundary modes at a pristine domain wall has remained challenging~\cite{Feldman2016}.

Subsequent unpublished measurements have allowed for investigation of a domain wall without nearby topographic complications. Preliminary results from this ongoing work include direct measurements of the valley occupation, LDOS, and spectral reconstruction in the vicinity of the domain boundary. The data are promising, and suggest that it may be possible to both directly address a Luttinger liquid with a local tunneling probe and to extract parameters relevant to characterizing the role of the domain wall in transport, which we discuss in the next subsection. Moreover, because $\mathcal{N}=6$ on the bismuth (111) surface, it offers the further possibility of exploring the same domain wall in regimes involving different numbers of counter-propagating quantum Hall edge modes at the boundary.  In the future, it may also be interesting to consider the properties of samples in an intermediate regime where there are only a few domains present, as these may provide routes to realizing `line junctions' and `Y-junctions' (for $\mathcal{N}\geq3$) of Luttinger liquids at nematic domain walls.

 \subsection{Domain Wall Transport: Theory and Experiments}
 As noted above, domain walls play a central role in elucidating detailed features of transport in multidomain samples. When gapped (localized) by valley-mixing interactions (impurity scattering), the domain walls are weak Mott (localized) insulators, in which the charge (mobility) gap  at the domain wall represents the weakest barrier to charge transport. In the absence of an external uniform component to the random valley-Zeeman field (i.e., when the system has {\it statistical} valley symmetry), all nematic domains are equiprobable; in two dimensions, this means that the domain walls must  percolate through the sample. 
 
 The longitudinal conductivity in the multidomain regime will be dominated  by the percolating domain wall network, as it is able to effectively mediate charge transport across the sample. This domain-wall mediated transport has two characteristic experimental signatures: (i) it will be {\it isotropic} and (ii) exhibit activated behavior $\sigma_{xx} \propto e^{-\Delta_{{dw}}/T}$; here,    $\Delta_{{dw}}$ is a domain wall energy scale characteristic of the one-dimensional insulator at the domain wall. For reasonable assumptions of interaction and disorder strength, this is likely much smaller than the gap $\Delta_{{qp}}$ to charged quasiparticle-quasihole pair excitations in the bulk of a nematic domain~\cite{KumarSAPSLSDWs}.
  
 Next, consider how transport changes in response to an external strain-induced valley Zeeman field. Since the domain wall modes are not valley-polarized, $\Delta_{{dw}}$ is only weakly affected, but the valley Zeeman field will drive the domain wall network away from percolation, since it favors occupation of states in one valley over the other(s). As a consequence, the domain walls do not provide an uninterrupted path for electronic excitations to traverse the system. Instead, longitudinal transport is now mediated by a combination of tunneling across the nematic domains between domain walls, and transport along the walls themselves. For small nut nonzero $\Delta_v$, $\sigma_{xx}\propto e^{-\Delta^*/T}$ will still exhibit activated behavior, but now controlled by an energy scale $\Delta^*(\Delta_{{dw}}, \Delta_{{qp}}, \Delta_{v})$ that depends in a complicated manner on the precise details of the domain wall network and the single-particle and valley Zeeman energies. In contrast, as strain is increased, the sample will be dominated by a single domain, until for large $\Delta_v$,  $\sigma_{xx}( \propto e^{-\Delta_{sp}(\Delta_v)/T}$; since the dominant single-particle excitation in a monodomain sample is a valley flip, we anticipate that $\Delta_{sp}(\Delta_v)\sim \Delta_{sp}(0) + \Delta_v$. In this limit, we expect that longitudinal transport will now be anisotropic in accord with our discussions above. 
 
In order to make this appealing heuristic argument more precise, we should properly treat the transport in the percolating domain wall network. An intuitive picture that emerges is analogous to two copies of a Chalker-Coddington model~\cite{ChalkerCoddington} used to describe transitions between QH plateaus. Unfortunately,  unlike the exactly soluble doubled network models applied, e.g. to the spin quantum  Hall transition~\cite{SpinQHTransition}, these doubled Chalker-Coddington models do not have  SU(2) symmetry, likely precluding an exact analytical solution. Nevertheless, it may be possible to leverage network model techniques for efficient numerical simulations of transport in the domain wall network. Note also that the $\mathcal{N}>3$ case admits the possibility of a valley-charge-separated domain wall. Following the reasoning above, if the DW gap is interaction driven, we anticipate that there will be no QHE in transport for $\nu=1$, and only a quantized charge Hall effect for $\nu=2$, with no accompanying quantized thermal Hall effect~\cite{Nu2DWTheory}.

As we have mentioned above, valley polarization in AlAs quantum wells has been especially well-studied via electronic transport measurements, with a particular focus on transport as a function of uniaxial in-plane strain. Early measurements showed that gluing a sample onto a piezoelectric actuator allowed for in-situ tuning of strain~\cite{ShayeganStrain}, providing an experimental knob to control valley Zeeman splitting and therefore valley occupation. Tuning this applied strain along different crystallographic axes leads to a dramatic change in the longitudinal resistance  due to the different anisotropic orientations associated with each valley, yielding a sharp experimental signature of valley polarization~\cite{Shkolnikov2004}, and indicating that strain can be used to drive a sample into the single-domain regime.

Similar devices have also been used to explore the QH state that emerges in AlAs at $\nu$ = 1 even when no external strain is applied.  In particular, measurements of the energy gap at $\nu$ = 1 (extracted by fitting the temperature-dependent  longitudinal conductivity to activated behavior) show that its magnitude rises much more quickly as a function of applied strain than is expected from single-particle considerations~\cite{Shkolnikov2005}. These results have been attributed to the  formation of valley Skyrmion excitations~\cite{Shkolnikov2005}, in analogy to spin Skyrmions  previously studied in GaAs by tilted-fieldal~\cite{PhysRevLett.75.4290}) and NMR measurements~\cite{PhysRevLett.74.5112,Tycko1460}. We believe that an alternative explanation~\cite{abanin_nematic_2010} along the lines of the domain-wall mediated transport discussed at length above is also compelling. Recall that although in clean samples, a spontaneously valley-polarized nematic phase is expected in the absence of strain, as we have noted, due to the disorder present in real samples, multiple nematic domains are likely to be present. Within this picture, the increasing activation gap can be interpreted as a crossover from domain-wall mediated transport to transport dominated by the exchange-enhanced valley Zeeman gap of a mono-domain sample~\cite{KumarSAPSLSDWs}. Evidently, this scenario rests on the assumption that the zero-strain sample is sufficiently strongly disordered that it has multiple domains. This has been difficult to verify directly in AlAs samples to date, though  estimates based on the expected disorder strength suggest that this is indeed the case. A rough estimate of the domain size accounting only for the long-wavelength disorder due to donor impurities in a doping layer offset a distance $\xi_{d}$ from the samples suggests a  typical domain size comparable to $\xi_d$. However,  the challenge of estimating the scale of other effects (e.g., random strains) that are expected to be present  coupled with the exponential sensitivity of the Imry-Ma domain size in 2D means that the domain size could be significantly different from this estimate.  More detailed discussions and calculations may be found in~\cite{abanin_nematic_2010,KumarSAPSLSDWs}.

Ultimately, further experimental study is essential to settle the issue. Recent improvements in AlAs sample quality~\cite{Chung2018} may enable measurements of the strain-dependent energy gap of nematic states in the clean limit to distinguish between the two scenarios. In addition, materials such as Bi films, where it is possible to use STM to both directly image the formation of nematic domains as well as to perform independent spectroscopic measurements of $\Delta_{qp}$ and $\Delta_{{dw}}$ that can then be compared against transport data, are a promising route to pursue.

\section{Adding Complexity\label{sec:complexity}}

\subsection{Continuous valley symmetries, order by disorder, and order by doping}
Recall that when a pair of valleys have identical anisotropy, the energy functional is  symmetric with respect to $SU(2)$ rotations between the two valleys; thus, any valley-polarized state within this subspace will spontaneously break this $SU(2)$ symmetry, thereby leading to a `valley-wave' Goldstone mode. This continuous valley symmetry and the associated Goldstone excitations have significant implications for the sequence of broken-symmetry states in the ultraclean limit in systems with $\mathcal{N}=4$ or $6$ valleys.
 In both cases, as noted in Sec.~\ref{sec:VQHFM-th}, we find that at $\nu=2$ the Hartree-Fock energy is exactly degenerate between two classes of QHFMs: `Class I' states where both valleys in an  anisotropy pair are filled, and `Class II' states built by picking two different anisotropy pairs and forming $\nu=1$ states within each. Unless external perturbations due to strain or higher-order interaction effects break this degeneracy, the selection between these possibilities is predicted~\cite{QHFMOBD} to occur via a QH generalization of the `order-by-disorder' mechanism familiar from frustrated magnetism~\cite{villain_order_1980}. Observe that Class II states break two $SU(2)$ symmetries leading to two distinct branches of Goldstone modes. Perhaps less obviously, a careful symmetry analysis shows that class I states do not break any symmetries and therefore host no Goldstone modes. Entropic state selection via thermal fluctuations will therefore prefer Class II states over Class I states as $T\rightarrow 0$. A similar conclusion can be inferred based on quantum fluctuations, via a mechanism dubbed `order by doping'~\cite{QHFMOBD}. The charge-ordered states  obtained on doping a Class II state are skyrmion crystals; these have a lower energy than the Wigner crystals that form around Class I states, and are hence favored by charge fluctuations around $\nu=2$. Therefore, thermal and quantum fluctuations both lead to the selection of the same class of states. Similar considerations may be applied at $\nu=3$ for the $\mathcal{N}=6$ case; here, absent Landau-level mixing, results for $\nu=4, 5$ follow by particle-hole symmetry about $\nu=6$.  For both $\mathcal{N} =4, 6$, the class II states at $\nu = \mathcal{N}/2$ break no spatial or other discrete symmetries and hence there is no finite-temperature transition into these states. A detailed analysis of the global phase diagram in these cases, developed in the context of Si(110)/(111) quantum wells, was performed in~\cite{QHFMOBD}.
 
\subsection{Ferroelectricity}
A recent theoretical analysis~\cite{Sodemann2017} predicts that valley QHFMs in materials whose Fermi pockets do not possess two-fold rotational symmetry are examples of a new class of QH ferroelectric phases. Specifically, they will not only exhibit nematic order, but will also possess an intrinsic in-plane dipole moment encoded in their wavefunctions. Experimental systems in which such physics can be realized include the surface states of materials in the family of the topological crystalline insulator Pb$_x$Sn$_{1-x}$Te, as well as individual valleys in bismuth. Single-valley polarization was recently reported in bismuth~\cite{Randeria2018ferro}, and high-resolution STM imaging of LL wavefunction interference in the vicinity of surface and sub-surface defects, in conjunction with detailed theoretical modeling, was used to identify the predicted ferroelectric ground state. In current experiments, the dipole moment is too small relative to the wavefunction extent to be directly imaged. Measurements of other materials or parameter spaces therefore represent particularly intriguing future directions, since this might provide the first opportunity to directly image QH ``valley-multiferroic'' behavior, as well as study domain walls between different ferroelectric polarizations. 

\subsection{Band structure warping and Lifshitz transitions}
Valley nematic phases can also emerge in materials as a result of Lifshitz transitions in the low-energy band structure. An example of such behavior was recently demonstrated in field-penetration capacitance measurements of Bernal (ABA) stacked trilayer graphene devices~\cite{Zibrov2018}. The trilayer graphene system is subject to trigonal warping, and detailed band structure calculations show that this results in three satellite valleys ('gullies') around the K and K' points in certain ranges of carrier density and perpendicular electric field. In a magnetic field, the threefold degeneracy around each valley was lifted, and incompressible states were experimentally observed at several intermediate filling factors. Numerical simulations suggest the broken-symmetry states have unequal occupation of each gully, which would constitute a new demonstration of a valley nematic. The high level of band structure tunability in response to external gating makes this system an exciting platform for future exploration.

\section{Summary and Future Directions \label{sec:summary}}
The combination of experimental measurements and theoretical understanding reported to date provide both motivation and a strong foundation for future study of quantum Hall valley nematics. In this review, we have highlighted several theoretical frameworks that can be used to understand such systems, and we summarize the current experimental status of various systems in Table \ref{table:CurrentStatus}. While substantial progress has been made, fully mapping out the phase diagram of quantum Hall valley nematics and investigating the exact role that domain walls play in sample properties are still in their infancy. Below, we discuss several future experiments of existing systems that are likely to yield new insight as well as two other related topics: fractional quantum Hall states and 3D systems.

\begin{center}
\begin{tabular}{ p{3cm} |p{1.3cm}| p{9.7cm} } \label{table:CurrentStatus}
 \bf{Material} & $\mathcal{N}$& \bf{Summary of experimental status}\\
 \hline
 AlAs & 2& Transport measurements of QHFM and nematicity, including anisotropic conductance, energy gaps, and strain-tuning of device behavior  \\ 
 \hline
 SnTe(111) & 3 & No measurements reported to date\\ \\  
 \hline
 Trilayer graphene (ABA) & 3 per K~point & Electronic compressibility measurements of QHFM states \\
 \hline
 Si(110) & 4& No measurements reported to date\\ \\  
 \hline
 Si(111) & 6 &  Transport measurements in QHFM regime \\
 \hline
 Bi(111) & 6 & Imaging of individual nematic and ferroelectric wavefunctions, preliminary investigation of domain walls \\
 \hline
 Pb$_x$Sn$_{1-x}$Te(100)  surfaces & 6 & QHFM and nematic behavior still to be experimentally explored \\
\end{tabular}
\end{center}

\subsection{Future Experiments} 
Future experimental efforts are likely to fall within several overarching categories. As described above, many open questions remain regarding the nature of nematic domain walls and their role in transport measurements. Both the bismuth and silicon materials systems are promising platforms to further investigate this topic because they offer direct access to scanning probe tips. Such measurements would offer insight into how the valley occupation evolves at the domain walls as well as the ability to investigate symmetry-protected topological edge modes and the degree to which they act as Luttinger liquids. Moreover, combining such measurements with externally applied valley Zeeman terms such as strain fields or in-plane magnetic fields would yield insight into how domain structure changes and how inter-valley mixing responds to these symmetry breaking terms. For example, an STM tip could track domain wall motion and changes in the spectrum near the domain wall as a function of such external perturbations and in response to variations in sample disorder.

A second major area that remains relatively unexplored in experiment is the full phase diagram of quantum Hall nematics as a function of temperature and disorder. As discussed above, the degree of disorder can influence not only the nature of the ground state, but also the detailed mechanism of electronic transport through devices. Temperature serves as an additional tuning knob within this phase diagram, both to control the nature of the broken symmetry and also the phase transition to an isotropic fluid.

Finally, investigating these phenomena in new materials and new parameter spaces represents a promising direction to pursue. For example, exploring the degree to which nematic behavior occurs in the fractional quantum Hall regime with a local STM probe would be exciting (see also the discussion below). Other materials remain to be studied, including the SnTe family, whose (111) surface has $\mathcal{N}=3$ and can be tuned with applied in-plane magnetic field, and whose (100) surface could provide a more favorable system to directly visualize the dipole moment of a quantum Hall ferroelectric wavefunction.

\subsection{Beyond Integers}

To date, the majority of experimental efforts have addressed nematic states within the IQH regime of anisotropic multi-valley systems. An intriguing area of inquiry is the degree to which similar rotational symmetry breaking emerges in the FQH regime. Of particular interest is the question of whether electronic band anisotropy can also influence FQH states, whose formation is driven by strong electron-electron interactions. We discuss several pioneering experiments that have probed this key question in GaAs and AlAs quantum wells below.

Anisotropy was first studied in the context of composite fermions (CFs), which can be understood as FQH quasiparticles consisting of an electron `bound' to an even number of flux quanta~\cite{Jain:1989p1,Jain2007}. In the CF model, these compound particles experience a reduced magnetic field, and the hierarchy of almost all FQH states can be explained as the IQHE of emergent CFs. To explore whether CFs inherit the band anisotropy of electrons, the dependence of sample resistance on strain was measured in AlAs at even-denominator filling factors, where a Fermi sea of CFs is present and the CFs experience no effective field ~\cite{Gokmen2010_NPhys}. These measurements show similar evolution of resistance at $\nu$ = 1/2 and at zero magnetic field, with opposite behavior for $\nu$ = 3/2 at low carrier densities. Such measurements were interpreted in terms of valley occupation at the Fermi level and are consistent with anisotropic CFs possessing qualitatively similar band anisotropy to electrons. In addition, strain has been shown to modulate the energy gaps of FQH states~\cite{Padmanabhan:2010p1} and induce phase transitions between states with different valley polarization~\cite{Bishop2007,Padmanabhan2009}. Subsequent measurements have quantitatively probed CF anisotropy induced by parallel magnetic field by measuring commensurability oscillations of electron~\cite{Kamburov2014} and hole~\cite{Kamburov2013, Mueed2015, Mueed2016, Jo2017} GaAs quantum wells. 

Future experiments may be able to shed further light on these questions. For example, the ability to image individual cyclotron orbits using a STM provided a direct visualization of nematic order for the first time in the IQH regime, and a natural question is whether similar techniques can be used to image FQH quasiparticles and their wavefunctions. Indeed, recent theoretical work suggests that it may be possible to image cyclotron orbits of CFs~\cite{Papic2017}. Such a technique could be applied to surface 2DEGs, such as Si surface states, where a robust FQHE has already been observed at experimentally accessible temperatures and magnetic fields. In addition, while the carrier density at the surface of bulk bismuth makes it impossible to experimentally address low LL orbital indices, doped samples or gated thin films may be used to bring the Fermi level to the lowest LL, where interaction effects are expected to be most pronounced. These potential experiments as well as continued study of the AlAs and GaAs systems represent exciting future directions to pursue. In a different vein, we may also ask whether the complex of phenomena associated with nematicity persist into the FQH regime, and in particular, if  the domain wall physics in this case allows access to new and unusual states at domain boundaries. 

On the theoretical front, the exploration of multi-valley systems motivate the construction of more intricate topological NLSMs that can properly capture the interplay of topological defects, fractionalization, and the emergence of gapless degrees of freedom in incompressible FQH valley nematics.  Some early numerical evidence has shown that the connection between the microscopic and emergent anisotropies may be significantly complicated away from integer filling, and may shed light on more subtle phenomena such as the underlying `quantum metric' of FQH states~\cite{PhysRevLett.107.116801,PhysRevB.85.165318,PhysRevLett.119.146602,PhysRevX.7.041032}.

\begin{figure}
\includegraphics[width=\columnwidth]{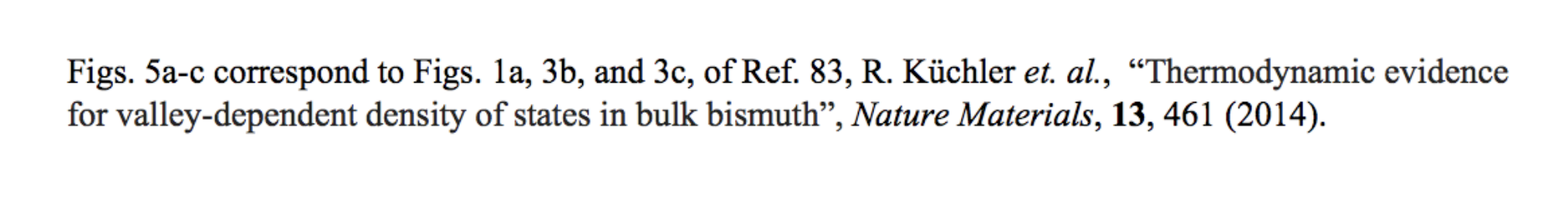}
\caption{Magnetostriction data in 3D Bi. (a) Schematic of apparatus used for measurement. (b) Magnetostriction data showing an imbalance of electron pockets e1 and e3 for positive and negative angles $\theta$, respectively, of the applied field relative to the trigonal axis. (c) Imbalance of the filling for valleys e1 and e3 for opposite signs of $\theta$. The dispersion is unchanged except for a relative shift $\delta$ in energy.
\label{fig:BehniaMagnetostriction}}
\end{figure}

\subsection{Beyond 2d}

Finally, we close by addressing the possibility that similar nematic quantum Hall physics can emerge even in bulk 3D samples. Bulk bismuth is a nearly compensated semimetal with a single anisotropic hole pocket along the trigonal axis, as well as three cigar-shaped electron pockets that are tilted just out of the binary-bisectrix plane and are rotated by 120 degrees relative to one another~\cite{Edelman1976}. A multitude of techniques have shown that valley polarization and rotational symmetry breaking emerge in bulk bismuth at low temperature and high magnetic field. These include measurements of angle-dependent magnetoresistance~\cite{Zhu2012, Collaudin2015}, torque magnetometry~\cite{Li2008}, and magnetostriction~\cite{Kuchler2014}. An imbalance in magnetostriction amplitude for different electron valleys provides evidence of unequal valley populations at the Fermi level (Fig. \ref{fig:BehniaMagnetostriction}), while the observation of identical Landau spectra for each valley in the same experiment suggests that strain does not break the rotational symmetry. One proposed interpretation for this behavior is that nematic physics, similar to the 2D case, plays a role at the extremal energy of three-dimensional LLs, where the kinetic energy of carriers along the field axis vanishes. An alternative potential explanation involves a field-induced lattice distortion due to electron-phonon interaction~\cite{Mikitik2015}, and the observed valley imbalance in bulk bismuth at the Fermi level remains an active field of study.

\subsection*{Acknowledgements}
We thank D.A.~Abanin, K.~Agarwal, R.J.~Cava,  H.~Ding, A.~Gyenis, H.~Ji, S.A.~Kivelson, A.~Kumar, A.H.~MacDonald, M.T.~Randeria, S.L.~Sondhi, F.~Wu, and A.~Yazdani for collaborations on related work, R.~Bhatt, M.~Serbyn, I.~Sodemann and L.~Fu for discussions, and M.~Shayegan, M.~Padmanabhan, T.~Gokmen, B.~Kane, T.~Kott and M.~Grayson for sharing their experimental results over several fruitful conversations. We also thank M.~Shayegan for detailed comments on the manuscript. S.A.P. acknowledges support from NSF Grant DMR-1455366 at the University of California, Irvine during the early stages of writing this review, and the hospitality of the KITP, UCSB (which is supported by NSF grant PHY-1748958) where this review was completed.

\section*{References}
\bibliography{NematicValleys}

\end{document}